\documentclass[12pt]{article}
\newlength{\mytopmargin}
\newlength{\myleftmargin}
\usepackage{amsmath,amsthm,amssymb,amsbsy,epsfig,graphicx,color,multicol,subfigure}
\usepackage{array,calc}
\usepackage[enableskew]{youngtab}
\usepackage{rotating}
\usepackage{young,epic}
\usepackage{a4wide,bm}
\usepackage{url}
\usepackage{hyperref}

\newtheorem{prop}{Proposition}

\usepackage{amsmath,amsfonts,amssymb}

\usepackage{graphicx}

\begin{document}
%
%\begin{frontmatter}

\title{Probability of all eigenvalues real for products of standard Gaussian matrices}
\author{Peter J. Forrester }
\date{}
\maketitle
\noindent
\thanks{\small Department of Mathematics and Statistics, 
The University of Melbourne,
Victoria 3010, Australia email:  p.forrester@ms.unimelb.edu.au 
}

\begin{abstract}
With $\{X_i\}$ independent $N \times N$ standard Gaussian random matrices, the probability $p_{N,N}^{P_m}$ that all eigenvalues are real
for the matrix product $P_m = X_m X_{m-1} \cdots X_1$ is expressed in terms of an $N/2 \times N/2$ ($N$ even) and
$(N+1)/2 \times (N+1)/2$ ($N$ odd) determinant. The entries of the determinant are certain Meijer $G$-functions. In the case $m=2$
high precision computation indicates that the entries are rational multiples of $\pi^2$, with the denominator a power of 2, and that
to leading order in $N$ $p_{N,N}^{P_m}$ decays as $(\pi/4)^{N^2/2}$. We are able to show that for general $m$ and large $N$,  $p_{N,N}^{P_m} \sim
 b_m^{N^2}$ with an explicit $b_m$. An analytic demonstration   that
$p_{N,}^{P_m} \to 1$ as $m \to \infty$ is given.
\end{abstract}

\section{Introduction}
The topic of products of random matrices saw much progress in the two decades up to the mid 1980's. The
achievements of this era are summarized in the books \cite{BL85,CPV93}, as well as some articles in the
conference proceedings \cite{CKN86}. Interest in the topic seemed to die down somewhat for the subsequent
two decades, until in the last few years when a number of researchers, most with backgrounds in integrability properties of the eigenvalue
spectrum of large random matrices, have revisited this topic. This has seen the discovery of rich mathematical structures,
analogous to those known for certain classes of ensembles of single random matrices, for ensembles of products of
random matrices.

The products may be infinite --- in which case the quantity of interest is the Lyapunov spectrum \cite{Fo12, Ka13} --- or finite but
allowing for an arbitrary number \cite{BJW10,BJLNS11,PZ11,AB12,Ip13,AKW13,AS13,AIK13,KZ13,ARRS13,IK13}.  The random matrices being
multiplied typically have Gaussian entries, and an exception being one of the ensembles considered in \cite{ARRS13},
which involves products of sub-blocks of unitary random matrices. Thus the setting is different to that of random matrix
products as they occur in the study of disordered chains \cite{Sc57}, or the one-dimensional Anderson model
\cite{CTT10,CLTT13} where one typically encounters products of random $2 \times 2$ matrices, some elements of which are fixed.

The study of integrability properties of the spectrum of a product of two (rectangular) Gaussian matrices was first undertaken by
Osborn \cite{Os04} (see also \cite{KS10}) in the case of complex entries. This was then generalized by Akemann et al.~\cite{APS09} to the case of real entries.
Edelman et al.~\cite{EKS94} found a number of exact results associated with the eigenvalues of the product $Y^{-1} X$ for $X$, $Y$
$N \times N$ real Gaussian matrices. This study was subsequently extended by Forrester and Mays \cite{FM09}.

For $X$ a square real random matrix there is a (typically) non-zero probabiliy $p_{N,k}^X$ of their being exactly $k$ real
eigenvalues. Since the complex eigenvalues occur in complex conjugate pairs, this requires $k$ to have the same
parity as $N$. It was shown in \cite{FM09} that for the random matrix product $Y^{-1} X$ the probability that all eigenvalues
are real is given by
\begin{equation}\label{A}
p_{N,N}^{Y^{-1} X} = {(\Gamma((N+1)/2)^N \over G(N+1)},
\end{equation}
where $G(N+1) := \prod_{l=1}^{N-1} l!$, $(N \in \mathbb Z^+)$ is the Barnes-$G$ function. This has the large $N$ form \cite{BF11a}
\begin{equation}\label{4.1}
p_{N,N}^{Y^{-1} X} = N^{1/12} \Big ( {e \over 4} \Big )^{N^2 /4} e^{-\zeta'(-1) - 1/12} \Big ( 1 + O(N^{-1}) \Big ).
\end{equation}
In the work \cite{BF11a} the probability $p_{N,N}^{Y^{-1} X} $ was shown to have an interpretation relating to the ranks of certain
random tensors.

In the case of a single $N \times N$ real Gaussian matrix $X$, a result of Edelman \cite{Ed95} gives that
\begin{equation}\label{A1}
p_{N,N}^X = 2^{-N(N-1)/4}
\end{equation}
(see also \cite{KA05} and \cite[\S 15.10]{Fo10}). Both (\ref{4.1}) and (\ref{A1}) exhibit a leading order Gaussian decay, but with a
slower rate for $p_{N,N}^{Y^{-1} X} $, the corresponding bases being $(e/4)^{(1/4)}$ and $(1/2)^{(1/4)}$ for $Y^{-1}X$ and $X$
respectively. A recent numerical
study of Lakshminarayan \cite{La13}, motivated by a problem in quantum entanglement, has considered the real eigenvalues for the
matrix product
\begin{equation}\label{4.3}
P_m = X_m X_{m-1} \cdots X_1,
\end{equation}
where each $X_i$ is an $N \times N$ is a real standard Gaussian matrix. It was demonstrated that  for $N$ fixed the probability of all eigenvalues being
real increases as $m$ increases, and approaches 1. It is the purpose of the present  paper to investigate this phenomenon and related
questions analytically,
using theory developed in the very recent work \cite{ARRS13}, together with methods familiar from the study of $p_{N,k}^X $ in
\cite{FN07}.

\section{Real eigenvalues of products of real Gaussian matrices}
\subsection{Determinant formulas}
\setcounter{equation}{0}
With each entry of the $N \times N$ matrix $X_l$ an independent Gaussian, the probability measure associated with
$X_1,X_2,\dots, X_m$ is equal to 
\begin{equation}\label{w6}
\Big ( {1 \over 2 \pi }\Big )^{m N^2/2} \prod_{l=1}^m e^{ - {1 \over 2} {\rm Tr} \, X_l X_l^T} (d X_l),
\end{equation}
where, with $X_l := [ x_{j,k}^{(l)} ]_{j,k=1,\dots,m}$, $(dX_l) := \prod_{j,k =1}^m dx_{j,k}^{(l)}$.
In the case $m=1$, the key \cite{Ed95} to computing the corresponding eigenvalue distribution is the real Schur decomposition
\begin{equation}\label{11.Rb}
X = Q  R Q^T,
\end{equation}
where $ Q$ is an $N \times N$ orthogonal matrix with elements of the first row positive while
\begin{equation}\label{11.Rl}
R = \left [
\begin{array}{cccccc}
\lambda_1 & \cdots & R_{1,k} & R_{1,k+1} & \cdots & R_{1,m} \\
 & \ddots &  \vdots & \vdots & \cdots & \vdots \\
& & \lambda_k & R_{k,k+1} & \cdots & R_{k,m} \\
& & & Z_{k+1} & \cdots & R_{k+1,m} \\
& & &         & \ddots & \vdots \\
& & &         &        & Z_m \end{array} \right ].
\end{equation}
Here all elements not explicitly shown are zero, $m=(N+k)/2$, and $R_{ij}$ is of
size $p \times q$ with
$$
p \times q = \left \{
\begin{array}{ll} 1 \times 1 & {\rm if} \: \: i \le k, \: j \le k, \\
1 \times 2  & {\rm if} \: \: i \le k, \: j > k ,\\
2 \times 1  & {\rm if} \: \: i > k, \: j \le k, \\
2 \times 2 & {\rm if} \: \: i > k, \: j > k.
\end{array}
\right.
$$
The variables $\{ \lambda_j \}_{j=1,\dots,k}$  are the real eigenvalues of $X$, while each $Z_j$ is the $2 \times 2$
matrix 
\begin{equation}\label{11.Rl'}
Z_j = \left [ \begin{array}{cc} x_j & b_j \\ -c_j & x_j \end{array} \right ],
\qquad b_j, c_j > 0, 
\end{equation}
where $x_j$ is the real part of the $j$-th complex eigenvalue of $X$ and 
$y_j = \sqrt{b_jc_j}$ with $y_j$ the imaginary part of the $j$-th complex eigenvalue. In the special case $k=N$ the structure of (\ref{11.Rl}) thus
simplifies and we have
\begin{equation}\label{RT}
R = {\rm diag} ( \lambda_1,\dots, \lambda_N) + T,
\end{equation}
where $T$ is the strictly upper triangular $N \times N$ matrix with non-zero entries $t_{jk}$, $k > j$.

Following an idea of Osborn \cite{Os04} in the complex case with $m=2$, and extended in \cite{AB12,ARRS13} in the real case for general $m$,
for real square matrices $\{ X_i \}_{i=1,\dots,m}$ the real Schur decomposition (\ref{11.Rb}) admits the generalization
\begin{equation}\label{w7}
X_i = Q_i R_i Q_{i+1}^T \qquad (i=1,\dots,m)
\end{equation}
with $Q_{m+1} := Q_1$. Each $Q_i$ is an $N \times N$ orthogonal matrix with elements of the first row positive, and each $R_i$ has the structure
(\ref{RT}). Our task is to use (\ref{w7}) to change variables in (\ref{w6}) for the sector $k=N$ (all eigenvalues real) then to integrate over
all variables except the eigenvalues of $P_m$. This will give us $p_{N,N}^{P_m}$.

\begin{prop}
Let
\begin{equation}\label{Wa}
w_m(\lambda) = \Big ( {1 \over \sqrt{2 \pi}} \Big )^m
\int e^{- \sum_{l=1}^m x_l^2/2} \delta ( \lambda - \prod_{l=1}^m x_l) \, dx_1 \cdots dx_m
\end{equation}
and let $L$ denote the region
$$
\lambda_1 > \lambda_2 > \cdots > \lambda_N.
$$
We have
\begin{equation}\label{w7.1}
p_{N,N}^{P_m} =2^{-mN(N+1)/4} \Big ( \prod_{j=1}^N {1 \over \Gamma (j/2) } \Big )^m  \int_L \prod_{j=1}^N w_m(\lambda_l) 
\prod_{1 \le j < k \le N} (\lambda_j - \lambda_k) \, d\lambda_1 \cdots d \lambda_N.
\end{equation}
\end{prop}

\noindent
Proof. \quad We know from the working of \cite[Appendix A]{ARRS13} that in the case each $R_i$ in (\ref{11.Rb}) with appropriate subscripts
is given by (\ref{RT})
\begin{equation}\label{w9.1}
\prod_{l=1}^m (d X_l) = \prod_{1 \le j < k \le N} (\lambda_j - \lambda_k) \prod_{l=1}^m (d T_l) (Q_l^T D Q_l) d \lambda_l,
\end{equation}
where $(Q_l^T D Q_l)$ denotes the Haar measure on the space of orthogonal matrices with all entries in the first
row positive, and $\{ \lambda_j \}$ the eigenvalues of $P_m$. Furthermore substituting (\ref{RT}) for each $R_i$ in (\ref{11.Rb}) shows
\begin{equation}\label{w9.2}
\prod_{l=1}^m e^{ - {1 \over 2} {\rm Tr} \, X_l X_l^T}  = \prod_{l=1}^m e^{- {1 \over 2} \sum_{k=1}^N ( \lambda_k^{(l)})^2}
e^{- \sum_{j<k} ( r_{jk}^{(l)})^2}.
\end{equation}

Substituting (\ref{w9.1}) and (\ref{w9.2}) in (\ref{w6}) we see that the dependence on the eigenvalues and the auxiliary variables
factorizes. The integrations over the auxiliary variables can be carried out according to \cite[second displayed equation below (15.211)]{Fo10}
$$
\int (Q^T d Q) = \pi^{N(N+1)/4} \prod_{j=1}^N {1 \over \Gamma(j/2)}
$$
and
$$
\int e^{- \sum_{j<k} ( r_{jk}^{(l)})^2} \, (d T_l) = (2 \pi)^{N(N-1)/4}.
$$
The result (\ref{w7.1}) now follows by noting that $\lambda_k = \prod_{p=1}^m \lambda_k^{(p)}$. \hfill $\square$

\medskip
The weight function (\ref{Wa}) is precisely the distribution of the product of $m$ standard Gaussian random variables, to be denoted
${\rm N}^{(m)}[0,1]$. It is well known (see e.g.~\cite{AB12} and references therein) that this can be written as an inverse Mellin
transform
$$
w_m(\lambda) = {1 \over (2 \pi)^{m/2}} {1 \over 2 \pi i}
\int_{c - i \infty}^{c + i \infty}
\Big ( {\lambda^2 \over 2^m} \Big )^{-s} \Gamma^m(s) \, ds, \qquad c>0.
$$
Introducing the Meijer $G$-function
\begin{equation}\label{GG}
G_{p,q}^{m,n} \Big ( z \Big | {a_1,\dots, a_p \atop b_1,\dots, b_q} \Big )
=
{1 \over 2 \pi i}
\int_C {\prod_{j=1}^m \Gamma ( b_j - s) \prod_{j=1}^n \Gamma(1 - a_j + s) \over
\prod_{j=m+1}^q \Gamma (1-  b_j + s) \prod_{j=n+1}^p \Gamma(a_j - s) } z^s \, ds
\end{equation}
for an  appropriate contour $C$, this can be written
\begin{equation}\label{10.1}
w_m(\lambda) = {1 \over (2 \pi)^{m/2} } G_{0,m}^{m,0} \Big ( \Big ( { \lambda^2 \over 2^m} \Big ) \Big | {\underline{\qquad} \atop
0,\dots,0} \Big ).
\end{equation}
For $m=2$ we have the alternative expression in terms of the $K_0$ Bessel function
\begin{equation}\label{W1}
w_2(\lambda) = {1 \over \pi} K_0(|\lambda|).
\end{equation}

It is furthermore the case that $p_{N,N}^{P_m}$ can be written as a determinant.

\begin{prop}
We have
\begin{equation}\label{W11}
p_{N,N}^{P_m} =2^{-mN(N+1)/4} \Big ( \prod_{j=1}^N {1 \over \Gamma (j/2) } \Big )^m \det A,
\end{equation}
where for $N$ even
\begin{equation}\label{Ae}
A =  [ \alpha_{2j-1,2k} ]_{j,k=1,\dots,N/2},
\end{equation}
while
for $N$ odd
\begin{equation}\label{Ao}
A = \Big [ [ \alpha_{2j-1,2k} ]_{j=1,\dots,(N+1)/2 \atop k=1,\dots, (N-1)/2} \quad [\nu_{2j-1}]_{j=1,\dots,(N+1)/2}  \Big ].
\end{equation}
Here the matrix elements are specified by
\begin{align}\label{W11a}
\alpha_{j,k} & = \int_{-\infty}^\infty dx   \int_{-\infty}^\infty dy \, w_m(x) w_m(y) x^{j-1} y^{k-1} {\rm sgn} \, (y-x) \nonumber \\
& =: \langle x^{j-1} y^{k-1} {\rm sgn} \, (y-x) \rangle_{x,y \in {\rm N}^{(m)}[0,1]}
\end{align}
(recall we are using $ {\rm N}^{(m)}[0,1]$ to denote the distribution of the product of $m$ standard Gaussian random variables) and
\begin{equation}\label{W11b}
\nu_j =  \langle x^{j-1} \rangle_{x \in {\rm N}^{(m)}[0,1]}.
\end{equation}
\end{prop}

\noindent
Proof. \quad According to the method of integration over alternate variables (see e.g.~\cite[Prop.~6.3.4]{Fo10}), for $N$ even
$$
\int_L w_m(\lambda_1) \cdots  w_m(\lambda_N) \prod_{1 \le j < k \le N} ( \lambda_j - \lambda_k) =
{\rm Pf} \, [ \alpha_{j,k} ]_{j,k=1,\dots, N}.
$$
But for $w_m(x)$ even, $\alpha_{2j,2k} = \alpha_{2j-1,2k-1} = 0$, showing that the entries of the Pfaffian vanish in a
chequerboard pattern, allowing it to be written as the determinant (\ref{Ae}). 
The case $N$ odd follows by appropriately modifying the method of integration over alternate variables
\cite[Exercises 6.3 q.1]{Fo10}, and an analogous reduction of the resulting Pfaffian to a determinant of half the size. \hfill $\square$

\medskip
Our next task is to evaluate the matrix elements (\ref{W11a}) and (\ref{W11b}).
\begin{prop}
We have
\begin{equation}\label{11x}
 \alpha_{2j-1,2k} = {1 \over (2 \pi)^m} 2^{(j+k-1/2)m}
 G_{m+1,m+1}^{m+1,m} \Big ( 1 \Big | {5/2-j,\dots, 5/2-j,2 \atop 1, 1+k,\dots, 1+k} \Big )
\end{equation}
and
\begin{equation}\label{11y}
\nu_{2j-1} = \Big ( {1 \over \sqrt{2 \pi}} \Big )^m (\Gamma(j - {1 \over 2}))^m.
\end{equation}
Thus, for $N$ even
\begin{equation}\label{11a}
p_{N,N}^{P_m} = \Big ( \prod_{j=1}^N {1 \over \Gamma (j/2) } \Big )^m
\det \Big [  G_{m+1,m+1}^{m+1,m} \Big ( 1 \Big | {5/2-j,\dots, 5/2-j,2 \atop 1, 1+k,\dots, 1+k} \Big ) \Big ]_{j,k=1,\dots,N/2}
\end{equation}
while for $N$ odd
\begin{align}\label{11b}
p_{N,N}^{P_m} & = \Big ( \prod_{j=1}^N {1 \over \Gamma (j/2) } \Big )^m \nonumber \\
&
\times \det \Big [   \Big [  G_{m+1,m+1}^{m+1,m} \Big ( 1 \Big | {5/2-j,\dots, 5/2-j,2 \atop 1, 1+k,\dots, 1+k} \Big ) \Big ]_{j=1,\dots,(N+1)/2 \atop
k=1,\dots,(N+1)/2} \quad [( \Gamma(j - {1 \over 2}))^m  ]_{j=1,\dots,(N+1)/2} \Big].
\end{align}
\end{prop}

\noindent
Proof. \quad We first note that
\begin{equation}\label{2.26a}
 \alpha_{2j-1,2k} = 2 \langle x^{2j-2} y^{2k-1} \chi_{y > x} \rangle_{x,y \in  {\rm N}^{(m)}[0,1]},
\end{equation}
 where $\chi_J$ for $J$ true, $\chi_J = 0$ otherwise. Recalling (\ref{10.1}) and applying a simple change of variables shows
\begin{align*}
 \alpha_{2j-1,2k} = &  {1 \over (2 \pi)^m} 2^{(j+k-1/2)m} \nonumber \\
 & \times \int_0^\infty dx \, x^{j-3/2}  G_{0,m}^{m,0} \Big ( x  \Big | {\underline{\qquad} \atop
0,\dots,0} \Big ) \int_x^\infty dy \, y^{k-1}   G_{0,m}^{m,0} \Big ( y  \Big | {\underline{\qquad} \atop
0,\dots,0} \Big ).
\end{align*}

Use of computer algebra gives
$$
 \int_x^\infty dy \, y^{k-1}   G_{0,m}^{m,0} \Big ( y  \Big | {\underline{\qquad} \atop
0,\dots,0} \Big ) =   G_{1,m+1}^{m+1,0} \Big ( x  \Big |{1 \atop
0,k,\dots,k} \Big )
$$
and furthermore
\begin{align*}
\int_0^\infty dx &\, x^{j-3/2}  G_{0,m}^{m,0} \Big ( x  \Big | {\underline{\qquad} \atop
0,\dots,0} \Big )  G_{1,m+1}^{m+1,0} \Big ( x  \Big |{1 \atop
0,k,\dots,k} \Big ) \nonumber \\
&  = G_{m+1,m+1}^{m+1,m} \Big ( 1 \Big | {5/2-j,\dots, 5/2-j,2 \atop 1, 1+k,\dots, 1+k} \Big ),
\end{align*}
thus implying (\ref{11x}). The result (\ref{11y}) now follows by substituting (\ref{11x}) in (\ref{W11}) and
straightforward simplification.

It is furthermore the case that
$$
\nu_{2j-1} = 
\Big ( {1 \over \sqrt{2 \pi}}   \int_{-\infty}^\infty x^{2j-2} e^{- x^2/2} \, dx \Big )^m,
$$
which implies (\ref{11y}). Substituting this and (\ref{11y}) in (\ref{W11}) in the case (\ref{Ao}) and simplifying gives (\ref{11b}).
\hfill $\square$

\medskip
We remark that as well as occurring in the study of products of Gaussian random matrices, the Meijer-$G$ function also
occurs in random matrix theory in the study of the value distribution of determinants \cite{CM00, DL00} and the study of the
Cauchy two-matrix model \cite{BGS08,BGS12}. The limiting correlation kernels appearing in the latter works have been related
to that for the ensemble of generalized Wishart matrices $P_2^\dagger P_2$ in \cite{KZ13}.

\subsection{Evaluations}
Consider first the case $m=2$, and thus the product of two Gaussian matrices $X$, $Y$ say. Although we have no proof, high precision computer
calculations indicate that the Meijer $G$-functions in (\ref{11x}) are all rational multiples of $\pi^2$, and furthermore the denominator
of each  is a power of 2. For example, with $m=2$, $N=6$
\begin{equation}
 \Big [   G_{3,3}^{3,2} \Big ( 1 \Big | {5/2-j, 5/2-j,2 \atop 1, 1+k, 1+k} \Big ) \Big ]_{j,k=1,2,3} 
\mathop{=}\limits^{?} \pi^2 
\begin{bmatrix} \displaystyle{1 \over 2^2} &  \displaystyle{39 \over 2^5} & \displaystyle{10335 \over 2^{13}} \\[.4cm]
  \displaystyle{3 \over 2^5} &  \displaystyle{435 \over 2^{10}} & \displaystyle {72555 \over 2^{18}} \\[.4cm]
  \displaystyle{135 \over 2^{13}} &  \displaystyle{16695 \over 2^{18}} &  \displaystyle{15107715 \over 2^{25}} \end{bmatrix}.
\end{equation} 
Assuming the validity of these forms, use of (\ref{11a}) and (\ref{11b}) then give the exact values
\begin{align}\label{AA}
p_{2,2}^{XY} = {\pi \over 2^2}, \qquad & p_{3,3}^{XY} = {5 \pi \over 2^5} \nonumber \\
p_{4,4}^{XY} = {201\pi^2 \over 2^{13}}, \qquad & p_{5,5}^{XY} = {10013 \pi^2 \over 2^{20}} \nonumber \\
p_{6,6}^{XY} = {64011585 \pi^3 \over 2^{36}}, \qquad & p_{7,7}^{XY} = {31 625 532 537 \pi^3 \over 2^{47}} 
\end{align}

The first of these has been derived in the recent work \cite{La13} (see also \cite{SRL11} and Section \ref{2.3} below).
Note that the case $m=2$ is special in that the corresponding weight function has the $K_0$  Bessel function form (\ref{W1}).
We remark that the $K_0$ Bessel function also appears in other closed form evaluations in
mathematical physics, in particular relating to the two-dimensional Ising model \cite{BBC06,BBBG08}.
For an informative recent article relating to high precision computations and closed form evaluations we
refer to \cite{BC13}.

Analysis of the corresponding numerical values, extended to $N=25$ and formed into the ratio
$$
{ p_{2j-1,2j-1}^{XY}  p_{2j+1,2j+1}^{XY}  \over (p_{2j,2j}^{XY})^2 }
$$
indicates that for large $j$ this has the limit value $\pi/4$ and that for large $N$
\begin{equation}\label{pN2}
p_{N,N}^{XY} \sim (\pi / 4)^{N^2/2}.
\end{equation}
This is a faster decay rate than seen in (\ref{A1}) for $p_{N,N}^{Y^{-1} X}$ (compare bases $(\pi/4)^{1/2} \approx 0.886$ and
$(e/4)^{1/4} \approx 0.907$). In the next subsection an analytic derivation of (\ref{pN2}) will be given, as will the leading large $N$
form of $p_{N,N}^{P_m}$ for general $m$.

We now turn our attention to the case $N=2$. We read off from (\ref{11a}) that
\begin{equation}\label{p2}
p_{2,2}^{P_m} = {1 \over \pi^{m/2}}  G_{m+1,m+1}^{m+1,m} \Big ( 1 \Big | {3/2,\dots, 3/2,2 \atop 1, 2,\dots, 2} \Big ). 
\end{equation}
In Table \ref{ta1} we list the corresponding numerical values up to $m=10$. High precision computation was used, but
no evidence of special arithmetic structures was found for $m > 2$. Analysis of the ratio
$
(1 - p_{2,2}^{P_{m+1}})/(1 - p_{2,2}^{P_{m}})$ for successive $m$ up to 16 gave values $\approx$ 0.82 but slowly increasing
in the third decimal, so evidence for an exponential approach to unity was inconclusive.

\begin{table}
\begin{center}
\begin{tabular}{c||c}
$m$& $p_{2,2}^{P_m}$  \\\hline
2& 0.7853981634 \\
3 & 0.8357987202 \\
4 & 0.8716118625 \\
5 & 0.8982590645 \\
6 & 0.9186258752 \\
7 & 0.9344692620 \\
8 & 0.9469484311 \\
9 & 0.9568694180 \\
10 & 0.9648135032\\
\end{tabular}
\caption{\label{ta1} First ten decimal places of the probability $p_{2,2}^{P_m}$ that the
random matrix product $P_m = X_m X_{m-1} \cdots X_1$, with each $X_i$ a $2 \times 2$ 
standard Gaussian matrix, has all eigenvalues real.}
\end{center}
\end{table}

\subsection{Leading large $N$ form of $p_{N,N}^{P_m}$}
The known analytic result (\ref{A1}) for $m=1$ and the numerical conjecture (\ref{pN2}) for
$m=2$ both exhibit a Gaussian decay in $N$ for $p_{N,N}^{P_m}$, but with different
bases $b_m$, $b_1 < b_2$. 
It is possible to establish a Gaussian decay for each $m$, and furthermore to determine
$b_m$.

To begin, we know from \cite[eq.~(4.186)]{Fo10} that
$$
\log \prod_{j=1}^N \Gamma(j/2) \sim {N^2 \over 4} \log {N \over 2} - {3 \over 8} N^2 +
O( N \log N).
$$
Substituting this in (\ref{w7.1}) and changing variables $\lambda_l \mapsto (c_{m/2} N)^{m/2} \lambda_l$
shows
\begin{align}\label{AG}
\log p_{N,N}^{P_m} & \sim {3m N^2 \over 8} + {N^2 m \over 4} \log 2 c_{m/2}  \nonumber \\
& + \log \int_L \prod_{l=1}^N w_m((c_{m/2} N)^{m/2} \lambda_l) \prod_{1 \le j < k \le N}
(\lambda_j - \lambda_k) \, d \lambda_1 \cdots d \lambda_N \, + O(N \log N).
\end{align}
Furthermore, since $N$ is large, we can use knowledge of the large argument form of the
Meijer $G$-function in (\ref{10.1}) as given in e.g.~\cite[pg.~12]{AS13} to write
$$
w_m(N^{m/2} \lambda) = e^{- m c_{m/2} N |\lambda|^{2/m} /2+ O(\log N)},
$$
allowing us to replace the logarithm of the integral in the final line of (\ref{AG}) by
\begin{equation}\label{AG1}
I_{m,N} := \log \int_L \prod_{l=1}^N    e^{- m c_{m/2} N  |\lambda_l|^{2/m}/2 } \prod_{1 \le j < k \le N}
(\lambda_j - \lambda_k) \, d \lambda_1 \cdots d \lambda_N .
\end{equation}

It is known rigorously (see e.g.~\cite[eq.~(11.1.22)]{PS11}) that
\begin{equation}\label{IE}
\log I_{m,N} \sim N^2  \mathcal E,
\end{equation}
where with $\rho(x)$ denoting the scaled density of the Coulomb gas model implied by
(\ref{AG1}), supported on the single interval $(-a,a)$,
\begin{equation}\label{E3}
\mathcal E =  - \int_{-a}^a V(x) \rho(x) \, dx + {1 \over 2}  \int_{-a}^a dx_1 \, \rho(x_1) 
\int_{-a}^a dx_2 \,  \rho(x_2)  \log | x_1 - x_2|
\end{equation}
with $V(x) =  m c_{m/2} |x|^{2/m}/2$. Moreover $\rho(x)$ is such that (\ref{E3}) is
minimised, giving rise to the terminology `the equilibrium problem' with $\rho(x) \, dx$
the equilibrium measure, while the one-body
Boltzmann factor $e^{-V(x)}$ with $V$ proportional to $|x|^\alpha$
is referred to as the Freud weight.

\begin{prop}
Choose $c_{m/2}$ in (\ref{AG1}) so that $\rho(x)$ is supported on $(-1,1)$.
Then we have
\begin{equation}\label{E9}
\mathcal E = - {1 \over 2} \log 2 - {3 m \over 8}.
\end{equation}
\end{prop}

\noindent
{Proof.} \quad 
We know from  \cite{ST98} that choosing
\begin{equation}\label{E4}
 {m c_{m/2} \over 2} = {\Gamma(1/m) \Gamma(1/2) \over 2 \Gamma(1/m + 1/2)}
 \end{equation}
 implies  that  $\rho(x)$ is supported on $(-1,1)$, that it
 has the explicit value
 \begin{equation}\label{E5}
 \rho(x) = {1 \over m \pi} \int_{|x|}^1 {u^{1/m - 1} \over \sqrt{u^2 - t^2}} \, du,
  \end{equation}
and furthermore
 \begin{equation}\label{E6}
   \int_{-a}^a dx \, \rho(x)  \log | x - y| =  {m c_{m/2} \over 2} |y|^{2/m} - \log 2 - {m \over 2}.
  \end{equation}   
It follows from (\ref{E6}) substituted in (\ref{E3}) that
  \begin{align}\label{E7}
\mathcal E& = - {1 \over 2} \Big ( \log 2 + {m \over 2} \Big ) + {m c_{m/2} \over 4} \int_{-1}^1 \rho(x) |x|^{1/m}
\, dx \nonumber \\
& =  - {1 \over 2} \Big ( \log 2 + {3 m \over 4} \Big ) ,
\end{align}
where the second line follows upon use of the explicit form of $\rho(x)$ (\ref{E5}), (\ref{E4}),
and the Euler beta integral.
 \hfill $\square$
 
 \medskip
 Substituting (\ref{E9}) in (\ref{IE}), substituting the result of this in the second line of
 (\ref{AG}) and making use of (\ref{E4}) in the first shows that
  \begin{equation}\label{E10}
  \log p_{N,N}^{P_m} \sim   N^2 \Big ( - {1 \over 2}  \log 2  + { m \over 4} \log 
  \Big ( {\Gamma(1/m + 1) \Gamma(1/2) \over  \Gamma(1/m + 1/2)} \Big ) \Big ),
  \end{equation}
  or equivalently
   \begin{equation}\label{E11}
 p_{N,N}^{P_m} \mathop{\sim}\limits_{N \to \infty}
 b_m^{N^2}, \qquad   b_m = {1 \over \sqrt{2}} \Big (  {\Gamma(1/m + 1) \Gamma(1/2) \over  \Gamma(1/m + 1/2)} \Big )^{m/4}.
 \end{equation}
  Substituting $m=1$ we reclaim the leading large $N$ form implied by (\ref{A1}),
  $  p_{N,N}^{P_1} \sim 2^{-N^2/4}$, while setting $m=2$ we obtain the
  conjectured form (\ref{pN2}). We can check from (\ref{E10}) that $b_m$ in (\ref{E11})
  is an increasing function of $m$ which tends to unity as $m \to \infty$. This latter
  feature is consistent with all eigenvalues being real in this limit, a topic
  we now turn to from a different perspective in the case $N=2$, before returning to
  (\ref{11a}) and (\ref{11b}) to give a demonstration for general $N$.

\subsection{Alternative expression for $p_{2,2}^{P_m}$}\label{2.3}
As indicated, we conclude by deriving an alternative expression to (\ref{p2}) for $p_{2,2}^{P_m}$, which allows us to both read off the exact value of $p_{2,2}^{P_2}$, and to give
some insight into the phenomenon $p_{2,2}^{P_m} \to 1$ as $m \to \infty$
observed through simulation in \cite{La13}, and in our list of exact decimal values in Table  \ref{ta1}.
We then make use of  (\ref{11a}) and (\ref{11b}) to show that $p_{N,N}^{P_m} \to 1$ as $m \to \infty$
for general $N \ge 2$.

\begin{prop}
With the notation  $ {\rm N}^{(m)}[0,1]$ for the distribution of $m$ standard Gaussian random variables
as used above we have
\begin{equation}\label{12.0}
p_{2,2}^{P_m} = {1 \over 2} \Big ( \sqrt{\pi \over 2} \Big )^{m-1} \langle \sqrt{x^2 + y^2} \rangle_{x,y \in {\rm N}^{(m-1)}[0,1]}.
\end{equation}
\end{prop}

\noindent
Proof. \quad We seek a formula for $\alpha_{1,2}$ as defined by (\ref{W11a}) different to that in (\ref{11x}).
Now
\begin{equation}\label{12}
\alpha_{1,2} = \langle (y-x) \chi_{y > x} \rangle_{x,y \in {\rm N}^{(m)}[0,1]} =
\int_0^\infty ds  \, s \int_{-\infty}^\infty dx \,  w_m(x)  w_m(x+s).
\end{equation}
According to the definition (\ref{Wa}), upon carrying out the integration over $x_m$,
$$
w_m(x) = \Big ( {1 \over \sqrt{2 \pi}} \Big )^m \int_{-\infty}^\infty dx_1 \cdots  \int_{-\infty}^\infty dx_{m-1} \,
{1 \over |X^{(m-1)}|} e^{- \sum_{j=1}^{m-1} x_j^2/2} e^{- x^2/(2 (X^{(m-1)})^2)},
$$
where $X^{(m-1)} := \prod_{l=1}^{m-1} x_l$. Hence
\begin{align*}
&\int_{-\infty}^\infty dx \, w_m(x) w_m(x+s) \\
& = {1 \over \sqrt{2 \pi}}
\Big \langle e^{-s^2/(2(  (X^{(m-1)})^2 +  (Y^{(m-1)})^2))} \sqrt{ (X^{(m-1)})^2 +  (Y^{(m-1)})^2))} \Big \rangle_{x_l, y_l \in {\rm N}[0,1] \, (l=1,\dots,m-1)}.
\end{align*}

Substituting in (\ref{12}) allows the integration over $s$ to be carried out, showing that
$$
\alpha_{1,2} = {1 \over \sqrt{2 \pi}}\Big  \langle  \sqrt{ (X^{(m-1)})^2 +  (Y^{(m-1)})^2))} \Big \rangle_{x_l, y_l \in {\rm N}[0,1] \, (l=1,\dots,m-1)}.
$$
Substituting this in (\ref{W11}) with $m=2$ and recalling the definition of ${\rm N}^{(m-1)}[0,1]$ gives (\ref{12.0}). \hfill $\square$

\medskip
According to (\ref{12.0})
\begin{align}
p_{2,2}^{P_1} & = {1 \over \sqrt{2}} \label{14.1}\\
p_{2,2}^{P_2} & = {1 \over 2} \sqrt{\pi \over 2} \langle \sqrt{x^2 + y^2} \rangle_{x,y \in {\rm N}[0,1]} = {\pi \over 4}, \label{14.2}
\end{align}
where the second equality follows upon using polar coordinates. The result (\ref{14.1}) is the special case $N=2$ of Edelman's result (\ref{A1}),
while (\ref{14.2}) is the first of the results in (\ref{AA}), which as already remarked has been proved recently in \cite{La13} using different
integration methods.

Using (\ref{12.0}) we can get some insight into the $m \to \infty$ behaviour.  Thus one has that
$$
 \Big ( \sqrt{\pi \over 2} \Big )^{m-1}\langle  |x| \rangle_{x,y \in {\rm N}^{(m-1)}[0,1]} = 1, \qquad
  \Big ( \sqrt{\pi \over 2} \Big )^{m-1}\langle  x^2 \rangle_{x,y \in {\rm N}^{(m-1)}[0,1]} =   \Big ( \sqrt{\pi \over 2} \Big )^{m-1}
$$
telling us that the variance of the random variable $\prod_{p=1}^{m-1} |x_l|$ for $x_l \in {\rm N}[0,1]$ is exponentially larger than the mean.
Thus, as in vividly demonstrated by Monte Carlo simulation, to leading order the random variables $x$ and $y$ in (\ref{12.0}) are independent
(typically simulated values of $|x|$ and $|y|$ are close to zero, with occasional large values which contribute most to the final average occurring
independently).
This shows that for large $m$
$$
p_{2,2}^{P_m} \to {1 \over 2} \Big (  \Big ( \sqrt{\pi \over 2} \Big )^{m-1} \langle|x|  \rangle_{x\in {\rm N}^{(m-1)}[0,1]} +
 \Big ( \sqrt{\pi \over 2} \Big )^{m-1} \langle|y|  \rangle_{y\in {\rm N}^{(m-1)}[0,1]} \Big )  = 1,
$$
in agreement with the result of Table \ref{ta1} and the simulations of \cite{La13}.

In fact the formulas (\ref{11a}) and (\ref{11b}) can be used to show that more generally,
for any $N \ge 2$, $p_{N,N}^{P_m} \to 1$ as $m \to \infty$, in agreement with the
extended simulations of  \cite{La13}.

\begin{prop}
We have
\begin{equation}\label{UA}
\lim_{m \to \infty}
\Big ( {1 \over \Gamma(j-1/2) \Gamma(k)} \Big )^m
G_{m+1,m+1}^{m+1,m} \Big ( 1 \Big | {3/2-j,\dots, 3/2-j,1 \atop 0, k,\dots, k} \Big ) 
= \left \{ \begin{array}{ll}1, & j \le k \\ 0, & j > k \end{array} \right.
\end{equation}
and thus for $N \ge 2$ 
\begin{equation}\label{UB}
p_{N,N}^{P_m} \to 1 \qquad {\rm as} \quad m \to \infty.
\end{equation}
\end{prop}

\noindent
Proof. \quad Since from the definition (\ref{GG})
$$
G_{m+1,m+1}^{m+1,m} \Big ( 1 \Big | {3/2-j,\dots, 3/2-j,1 \atop 0, k,\dots, k} \Big ) 
= G_{m+1,m+1}^{m+1,m} \Big ( 1 \Big | {5/2-j,\dots, 5/2-j,2 \atop 0, k,\dots, k} \Big ) 
$$
we see from  (\ref{11a}) and (\ref{11b}) that (\ref{UB}) follows from (\ref{UA}), so it remains
to establish the latter.

Now  (\ref{GG}) gives
$$
G_{m+1,m+1}^{m+1,m} \Big ( 1 \Big | {3/2-j,\dots, 3/2-j,1 \atop 0, k,\dots, k} \Big ) =
- {1 \over 2 \pi i} \int_C {(\Gamma(k-s) \Gamma(j - {1 \over 2} + s))^m \over s} \, ds
$$
where $C$ can be taken to be a contour starting at $-i \infty$, passing through the real axis within
the interval $({1 \over 2} - j, 0)$ and finishing at $i \infty$. Changing variables
$s \to s/m$ we see that for large $m$
$$
 (\Gamma(k-{s \over m}) \Gamma(j - {1 \over 2} + {s \over m}))^m \to
 (\Gamma(k) \Gamma(j- {1 \over 2}))^m
 e^{-s (\Psi(k) - \Psi(j-1/2))},
 $$
 where $\Psi(z)$ denotes the digamma function. But for the contour $C$ running from $-i \infty$
 to $i \infty$ and passing to the left of the origin, and with $r$ real
 $$
 - {1 \over 2 \pi i} \int_C {e^{-s r} \over s} \, ds =
  \left \{ \begin{array}{ll}1, & r > 0  \\ 0, & r < 0\end{array} \right.,
  $$
  as is seen by closing the contours to the right $(r > 0)$, left $(r < 0)$. The result now
  follows since $\Psi(k) - \Psi(j-1/2)) > 0$ for $k \ge j$ while this quantity is less than zero for
  $k < j$. \hfill $\square$

\subsection*{Acknowledgements}
 This work was supported by the Australian Research Council. I thank Gernot Akemann for comments on the
 first draft, and Mario Kieburg for sending me a copy of \cite{IK13}.

%\bibliographystyle{amsplain}
%\bibliography{book1}

\providecommand{\bysame}{\leavevmode\hbox to3em{\hrulefill}\thinspace}
\providecommand{\MR}{\relax\ifhmode\unskip\space\fi MR }
\providecommand{\MRhref}[2]{%
  \href{http://www.ams.org/mathscinet-getitem?mr=#1}{#2}
}
\providecommand{\href}[2]{#2}

\end{document}